\documentclass[%
reprint,
superscriptaddress,
 amsmath,amssymb,
 aps,
]{revtex4-1}

\usepackage{graphicx}% Include figure files
\usepackage{dcolumn}% Align table columns on decimal point
\usepackage{bm}% bold math
\usepackage{hyperref}% add hypertext capabilities
\usepackage{amsmath}
\usepackage{comment}
\usepackage[export]{adjustbox}
\usepackage{float}

\begin{document}

\preprint{APS/123-QED}

\title{Refractory times for excitable dual state quantum dot laser neurons
}% Force line breaks with \\

\author{M. Dillane}
\affiliation{Department of Physics, University College Cork, Ireland}
\affiliation{Tyndall National Institute, University College Cork, Lee Maltings, Dyke Parade, Cork, Ireland}
\affiliation{Centre for Advanced Photonics and Process Analysis, Cork Institute of Technology, Cork, Ireland}

\author{E.A. Viktorov}
\affiliation{National Research University of Information Technologies, Mechanics and Optics, Saint Petersburg, Russia}

\author{B. Kelleher}
\affiliation{Department of Physics, University College Cork, Ireland}
\affiliation{Tyndall National Institute, University College Cork, Lee Maltings, Dyke Parade, Cork, Ireland}

\date{\today}% It is always \today, today,

\begin{abstract}
Excitable photonic systems show promise for ultrafast analog computation, several orders of magnitude faster than biological neurons. Optically injected quantum dot lasers display several excitable mechanisms with dual state quantum lasers recently emerging as true all or none excitable artificial neurons. For use in applications, deterministic triggering is necessary and this has previously been demonstrated in the literature. In this work we analyse the crucially important \emph{refractory time} for this dual state system, which defines the minimum possible time between distinct pulses in any excitable pulse train. Ultrashort times on the order of 1~ns are obtained suggesting potential use where ultrafast analog computing is desired.
\end{abstract}

\pacs{Valid PACS appear here}% PACS, the Physics and Astronomy
\maketitle

\section{Introduction}

Excitable spiking photonic systems have recently emerged as novel computational frameworks potentially leading to low energy, ultrafast brain inspired information processing \cite{shastri_crc, bhavin_review}. Originally observed in biological neurons \cite{izhikevich,izhikevich_2004}, it has since been observed in many different neuromimetic systems. In particular, it is found in the optically injected laser system \cite{wiecz_prl, goulding_prl, guy_srl, wells_v_dots, hurtado_vcsel}. When an excitable system is subject to perturbations it can respond in two qualitative manners. For small enough perturbations the system returns to steady state along a short phase space trajectory while for sufficiently large perturbations - those above the excitable threshold - there is a large phase space trajectory in the return to steady state, typically known as a pulse or a spike. In biological neurons this manifests as a voltage spike across the membrane of the cell while it laser systems it typically takes the form of an intensity spike with an associated characteristic electric field phase evolution that depends on the particulars of the configuration.

In an excitable system, the faster the ``neuron" can process data the better, and so, one might hope to deterministically trigger pulses with high repetition rates. However, there is an intrinsic limit to the allowed durations between excitations. After a pulse is fired the system takes some time to recover as it returns to steady state. Only after this time has passed can another pulse be triggered. This minimum time is known as the absolute refractory time. For times shorter than the absolute refractory time it is impossible to trigger a second pulse. Following the absolute refractory time it is possible to trigger a second pulse, but the pulse amplitude is inhibited for times that are shorter than the so-called relative refractory time. After this time has passed, a full excitable pulse can be fired again. Therefore, the shortest time between full excitable pulses is the relative refractory time.

Considering the number of investigations of excitability in the optical injection configuration, there are relatively few analyses of the refractory times. Most relevant to the work presented here is the work in \cite{garbin2017refractory} where the absolute refractory time in an optically injected vertical cavity semiconductor laser was investigated both experimentally and theoretically, with phase perturbations used to excite the pulses. Previous studies have also found the absolute and relative refractory times for a semiconductor micropillar laser with a saturable absorber \cite{selmi2014relative}. In this work we analyse the refractory times for an optically injected, dual state InAs based quantum dot (QD) laser. QD lasers are unique among semiconductor lasers in that they can emit from two (or more) distinct energy states \cite{Markus2003SimultaneousLasers, jenya_apl, rohm_jqe}. The device used in this work can emit from the ground state (GS) and the first excited state (ES) depending on the pump current. When free-running we pump it so that it emits only from the ES. It is then optically injected by an external tunable laser emitting close to the frequency of the GS. For sufficiently high injection strengths and low detuning (the frequency of the tunable laser minus that of the GS of the QD laser) the optical injection can completely suppress the ES emission and lasing from the GS only can be obtained. For relatively low injection strengths and negative detuning values a dynamical region of periodic deep GS dropouts and accompanying short ES intensity pulses is obtained. Progressively moving towards zero detuning the periodic behaviour is replaced by randomly spaced dropouts/pulses \cite{interplay} and eventually a stable, phase-locked output from the GS only. The random dropouts/pulses are stochastic excitable events. Importantly, these excitable responses can also be triggered deterministically via short phase perturbations \cite{md_ol} in the phase-locked region. An intriguing feature of this excitable regime is that sharp excitable thresholds exist for both clockwise and anticlockwise phase perturbations. This is in direct contrast to the excitable regime found for negative detuning at low injection strengths with conventional semiconductor lasers and even with single state QD lasers, where only anticlockwise perturbations yield an excitable response. For the dual state lasers, the two thresholds are asymmetric with the clockwise threshold typically larger than the anticlockwise one. Thus, the study of the refractory behaviour of the dual state laser system requires separate analyses of the two directions. Finally, unlike the vast majority of neuromorphic systems, the excitable ES pulses are true all-or-none responses, as also shown in \cite{md_ol}.

\begin{figure}[t]
\centering
\includegraphics[width=\linewidth]{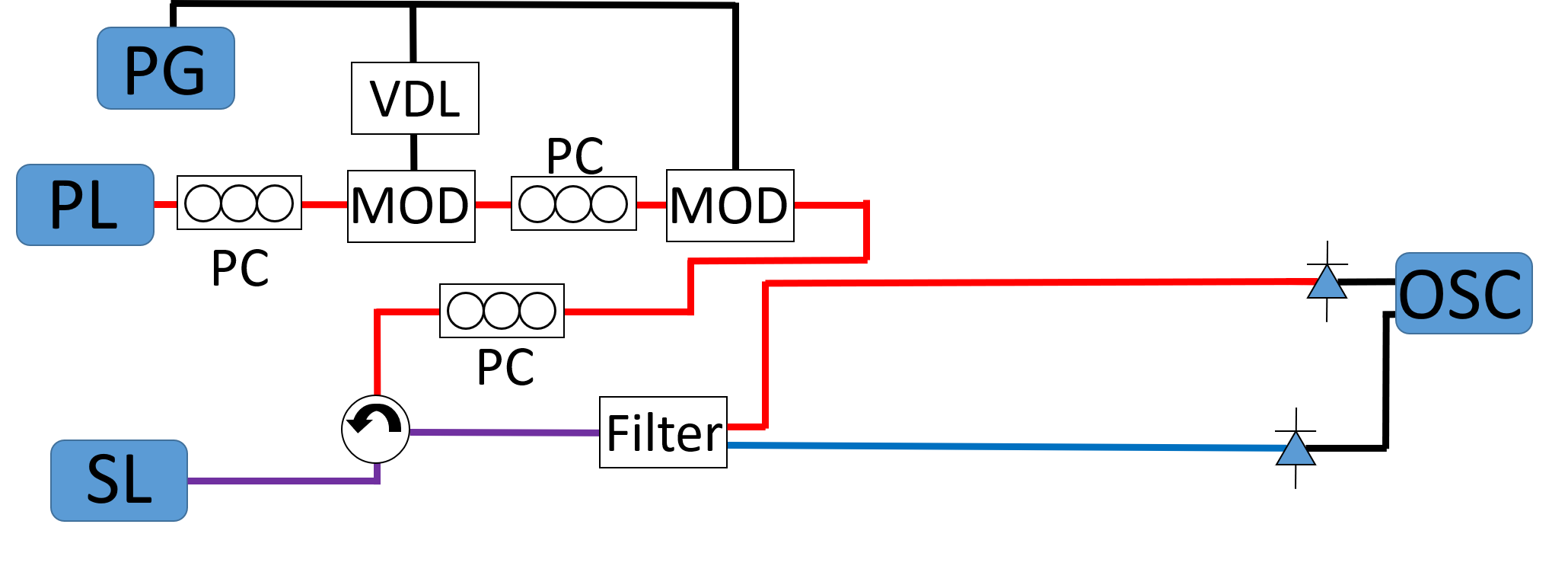}
\caption{The secondary laser (SL) is a QD laser and the primary laser (PL) is a tunable laser source (TLS). The LiNbO$_3$ phase modulators (MOD) are driven by a pulse generator (PG). A 50:50 power splitter sends a similar signal to each modulator. A variable electrical delay line (VDL) is added to adjust the time between two perturbations. A polarisation controller (PC) is used to maximise coupling. The light emitted from the SL is sent into the circulator and then to a filter where the ES and GS light are separated and are sent directly to 12~GHz detectors connected to an oscilloscope (OSC). The red lines represent light at approximately 1300~nm close to the GS emission, the blue represents ES light only and the purple is both GS and ES. The black lines are high speed electrical cables.}
\centering
\label{exp_ref_small}
\end{figure}

\section{Experimental setup}

Our perturbations take the form of square phase pulses generated by square voltage pulses sent to phase modulators that quickly change the phase of the light from the tunable laser \cite{md_ol}. To measure the refractory time, two perturbations are required and so the voltage pulse is split using a 50/50 splitter with one half going to one phase modulator and the other half to a second. A variable delay line allows us to change the time between the two resulting phase perturbations. Figure \ref{exp_ref_small} shows a schematic of the experiment. The magnitude of the detuning is set to the maximum value possible that allows for stable phase locking, without the appearance of any stochastic excitable pulses. Initially we set the magnitude of the perturbations so that the first perturbation alone always triggers a pulse. The time between the perturbations is then varied and the responses of the system measured and analysed.

\section{Anticlockwise perturbations}

Since our perturbations arise from square voltage pulses there is a clockwise perturbation and an anticlockwise perturbation in each square. The clockwise part comes from the rise of each square and the anticlockwise part from the fall of each square. The rise time of $\sim$200~ps is much faster than the fall time of $\sim$470~ps. The duration of the pulse is set to 5~ns, much longer than the intrinsic timescales and likely refractory times. For the control parameters used here, the threshold magnitude for anticlockwise perturbations is 2~rad and for clockwise is -3.5~rad. The perturbation strength is initially set to approximately $\pi$ rad which has a pulse triggering efficiency of 100\% for the anticlockwise part.

\begin{figure}[t]
\centering
\includegraphics[width=\linewidth]{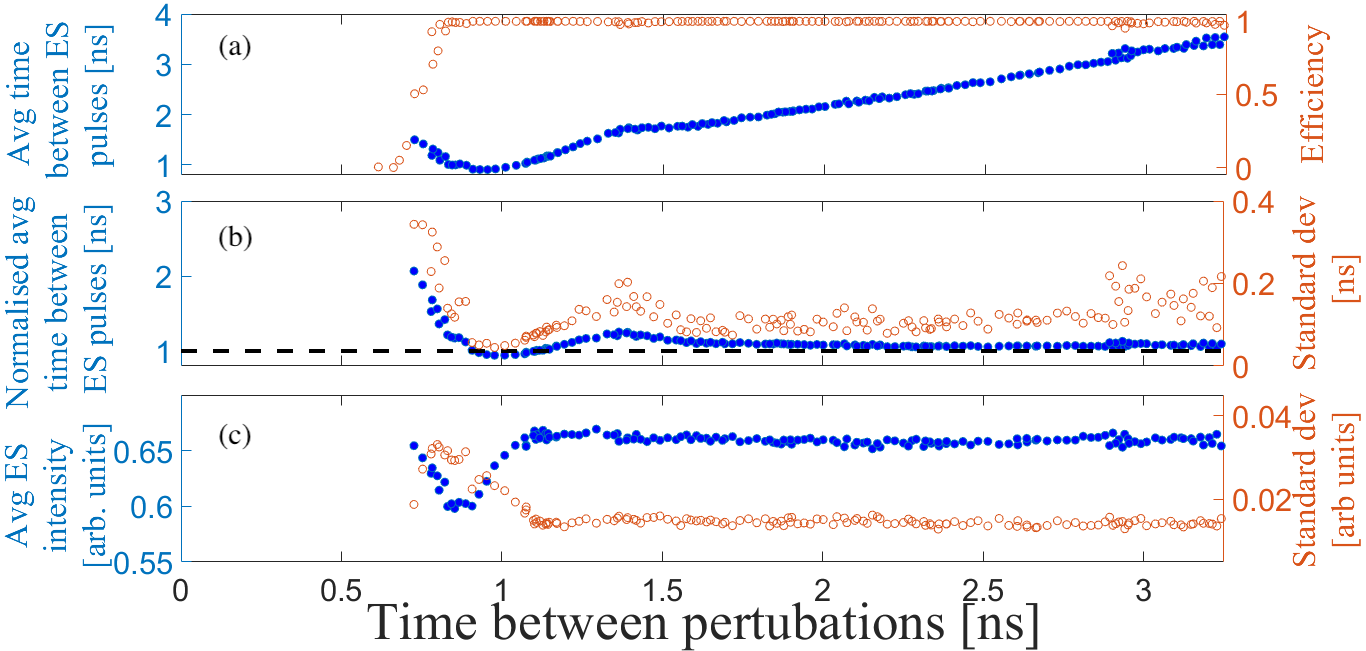}
\caption{Anticlockwise perturbations are used in this Figure. (a) The blue dots show the average time between two ES pulses after two perturbations were applied vs the time between two perturbations. Each data point is the average of 1000 sets of perturbations. The first perturbation is always successful. The absolute refractory time is 0.62~ns. The red dots show the efficiency of the second perturbation. (b) The blue dots show the normalised average times between two ES pulses after two perturbations were applied vs the time between perturbations. The average is of 1000 sets of perturbations. The red line shows the standard deviation of the times between two ES pulses. (c) The blue dots show the average peak intensity of the second ES pulse vs the time between perturbations. The red dots show the standard deviation of the peak intensities vs the time between perturbations.}
\centering
\label{optimal}
\end{figure}

Figure \ref{optimal} (a) shows the efficiency of the triggering of a second pulse versus the time between the two perturbations (in the cases where both perturbations successfully triggered pulses). If the perturbation separation is shorter than 0.62~ns, then a second pulse is never fired. That is, there is 0\% efficiency for times less than 0.62~ns and we identify this time as the absolute refractory time. When the times between perturbations are increased to 0.8~ns the efficiency is 100\%.

Figure \ref{optimal} (b) shows the time between the two pulses normalised by the time between the perturbations and also shows the standard deviation of times between ES pulses. Close to the absolute refractory time the average time between the pulses can be quite long - more than twice the time between the perturbations as can be seen in the figure. Furthermore, the standard deviation is very large, highlighting the strong influence of noise on the delay time of the second pulse when the time between the perturbations is close to the refractory time. Between 0.8~ns and 1.2~ns the normalised time between two ES pulses goes below 1. This means that the time between pulses is less than the time between perturbations. The standard deviation of times between ES pulses is smallest in this region. Conversely, between 1.2 and 1.6~ns the normalised time between two ES pulses increases beyond 1 and the standard deviation of times between the two ES pulses increases again to a local maximum. This also corresponds to a small decrease in efficiency shown in Figure \ref{optimal} (a) to 98\% at 1.36~ns. We ascribe this behaviour to the trajectory followed by the system as it relaxes back to the steady state via the highly damped GS relaxation oscillations associated with QD lasers \cite{ludge_damping,switching}. This manifests mostly as an overshoot of the intensity and phase following each dropout. That is, following the dropout, the intensity increases to a value above the steady state value before relaxing back to the steady state value. Similarly with the phase. As a result, during the return to steady state the trajectory can take the system closer to or further from the separatrix that must be passed to excite a pulse, than when settled in the steady state. Thus, for the region where na\"{i}vely it might seem like the system is preempting a pulse, this is of course not the case and the second pulse is indeed triggered by the second perturbation. Instead, we interpret this as showing that the system is relatively closer to the excitability separatrix when the second perturbation arrives, and hence it doesn't take as much of the perturbation's rise time to excite a pulse. Thus, the perturbation acts like an effectively larger perturbation. This also means that the perturbation moves the system well past the separatrix and so the effect of noise is diminished and the standard deviation is smaller as found.

On the other hand, where the normalised time between pulses is above 1, the system is further from the separatrix than the steady state case and so the amplitude of the second perturbation is not large enough to perturb the system far enough beyond the separatrix to fully negate the effects of noise. There is thus a subsequent large distribution of escape times and an increase in the probability of failure. When the time between perturbations goes beyond approximately 1.6~ns in Figure \ref{optimal} (b) the normalised time between two ES pulses remains constant at 1, and the standard deviation remains fixed at 1.5. We thus identify 1.6~ns as the relative refractory time.

Figure \ref{optimal} (c) shows the average peak intensity of the second ES pulse versus the time between perturbations and the standard deviation of the peak intensities versus the time between perturbations. Similar to (b) there is a large variation in behaviours from the absolute refractory time up to approximately 1.6~ns, from which point on the amplitudes of the pulses and dropouts return to their single perturbation values, agreeing with our earlier identification of 1.6~ns as the relative refractory time.

\begin{figure}[t]
\centering
\includegraphics[width=\linewidth]{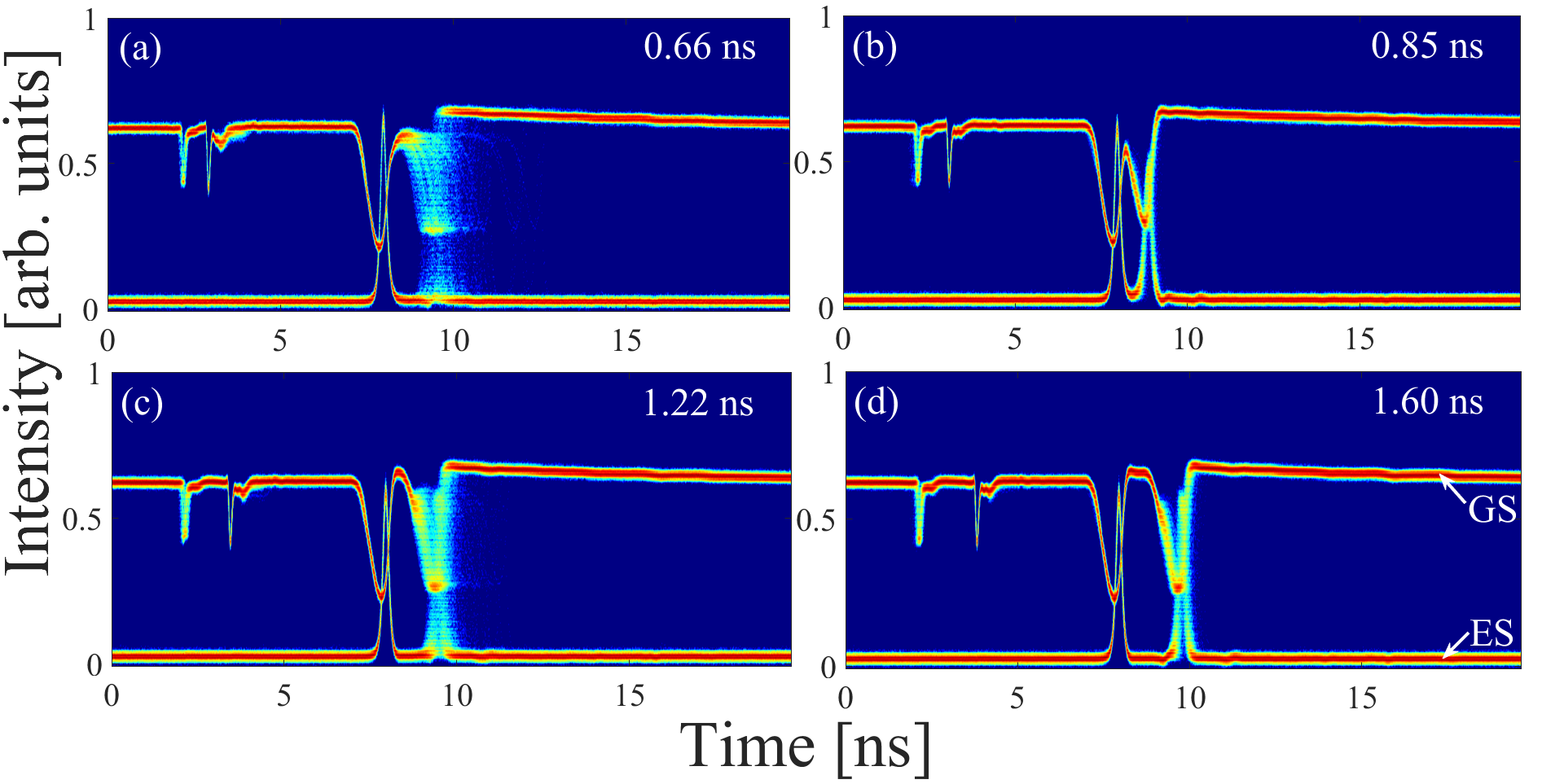}
\caption{The two small GS negative spikes at approximately 2~ns in each panel are the locations of the subthreshold clockwise perturbations. (a) Timetraces with perturbation separation of 0.66~ns: close to the absolute refractory time. The second pulse can have long escape times. (b) Perturbations separated by 0.85~ns. There is a narrow distribution for the second pulse. (c) Perturbations separated by 1.22~ns. There is a large distribution of delay times for the second pulse. (d) Perturbations separated by 1.6~ns. The second pulse has almost identical characteristics to the first. }
\centering
\label{all_traces}
\end{figure}

Figure \ref{all_traces} is complementary to Fig. \ref{optimal} and shows the responses from several different perturbation separations with 1000 different runs of the experiment overlaid in each case. The small variation in delay times for the first pulse is clear from the very tight overlap seen in each panel of Figure \ref{all_traces}. The two dips in the GS intensity seen early in each panel correspond to the clockwise part of each square pulse, which allows for a convenient measurement of the time between perturbations. In Fig. \ref{all_traces} (a), the time between perturbations is 0.66~ns, slightly higher than the absolute refractory time. The variation in delay times for the second pulse is clear from the jittered, spread out set of pulses. Figure \ref{all_traces} (b) shows the behaviour for a separation of 0.85~ns between the perturbations. One can see that the GS intensity does not return to its steady state value before the second dropout is triggered. The small standard deviation in the timing of the second pulse is clear. In (c) the time between perturbations is 1.22~ns and in contrast to (b) there is a large standard deviation in the second pulse timing. Finally, in (d) the characteristics of the second pulse have almost returned fully to those of the first pulse.

\section{Clockwise perturbations}
We now analyse the influence of the clockwise perturbations. As already mentioned, such perturbations do not yield excitable responses in conventional laser systems and so there is no associated threshold in the conventional case, by definition. The asymmetry in the threshold perturbation strength means we must increase the magnitude of the perturbations. To this end, we use clockwise perturbations of magnitude 5.5~rad at which strength, single clockwise perturbations (the rising parts of the voltage pulse) excited pulses with 100\% efficiency. However, something novel arises when attempting to measure the refractory times by including the second perturbation. In order to measure the refractory time, a pre-requisite is that the first perturbation must trigger a pulse. This turns out to be a non-trivial situation for clockwise perturbations with the applied rise time. To guarantee that the first perturbation triggers a pulse, we find that the two perturbations have to be at least 1~ns apart. For times less than 1~ns, an inhibitory integration mechanism is found as discussed in \cite{LIF_PPI}. We thus consider the case where the perturbations are at least 1~ns apart in this work.

\begin{figure}[h]
\centering
\includegraphics[width=\linewidth]{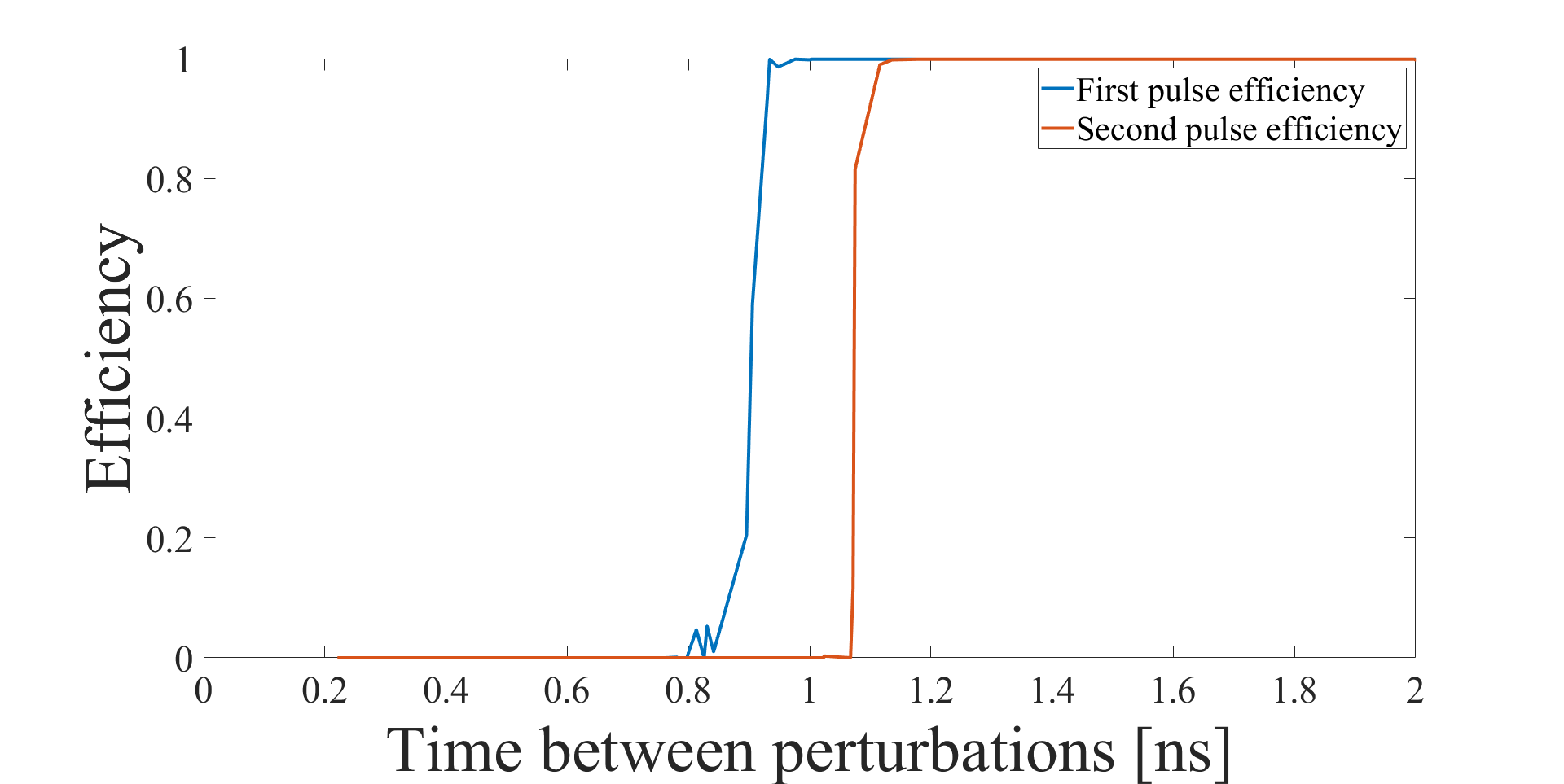}
\caption{Efficiency curves for the first clockwise perturbation (blue) and the second clockwise perturbation(red) given that the first perturbation has already triggered a pulse.}
\centering
\label{up_effs}
\end{figure}

A pair of efficiency curves are plotted in Figure \ref{up_effs} showing the efficiency curve for both perturbations. If the second perturbation arrives at least 1~ns after the first, there is 100\% probability that the first perturbation will trigger a pulse. When the two perturbations are separated by 1.02~ns, there is a finite (~0.003\%) probability that the second perturbation will trigger a pulse. Thus, we identify 1.02~ns as the absolute refractory time. The influence of the inhibitory integration mechanism means that the lack of a second pulse is not necessarily due to the usual refractory time mechanisms. We take a pragmatic view in this work: since one must have at least 1.02~ns between perturbations in order to generate two pulses, we define this as the absolute refractory time, while acknowledging that the system is more complex than the conventional case.

\begin{figure}[t]
\centering
\includegraphics[width=\linewidth]{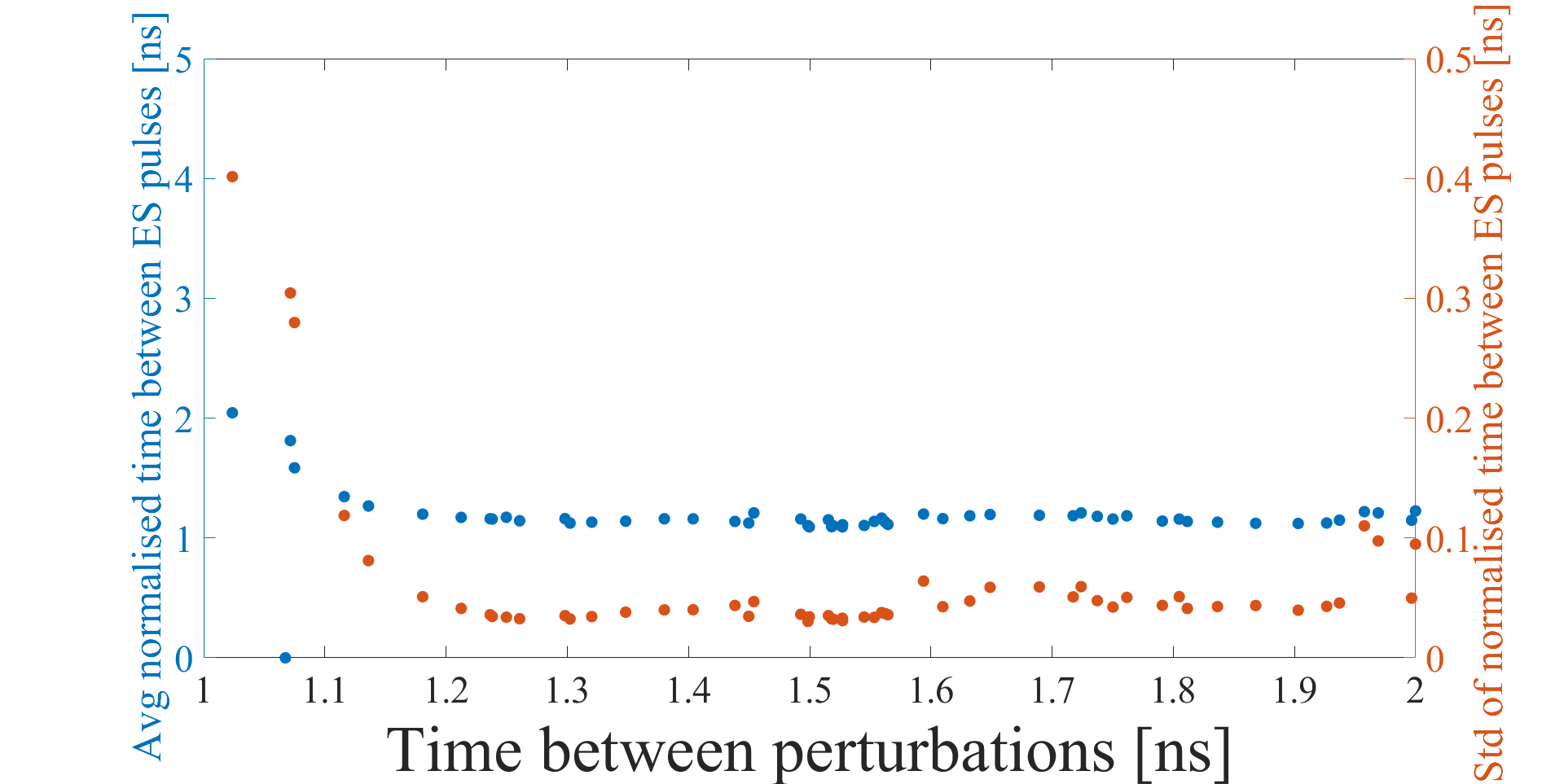}
\caption{The blue curve shows the average time between two ES pulses after two clockwise perturbations are applied. The red curve shows the standard deviation of the times between two ES pulses.}
\centering
\label{up_ref_trace}
\end{figure}

When the time between perturbations is increased to 1.18~ns the second perturbation is 100\% successful in triggering a pulse.
The shape of the normalised time between two ES pulses curve shown in Figure \ref{up_ref_trace} is similar to the anticlockwise case with a large variation in behaviour from the absolute refractory time up to the relative refractory time, which we identify as 1.25~ns. The average time between ES pulses is large close to the absolute refractory time, after which it drops, levelling off to match the perturbation separation at the relative refractory time.

\section{Large anticlockwise perturbations}

Because of the nature of our perturbations we have an accompanying anticlockwise perturbation of 5.5~rad with each clockwise one. Intuitively it is clear that refractory times are not independent of the perturbation amplitude. Consider perturbations slightly above the excitable threshold. Following the initial perturbation the system follows the deterministic trajectory back to the steady state. Before the absolute refractory time has passed the second perturbation cannot trigger a pulse. If the magnitude of the perturbation is now increased then the excitable pulse resulting from the first perturbation is almost unchanged but there is an increased likelihood of triggering the second pulse as the larger perturbation might force the system from the trajectory. Thus one can expect the absolute refractory time to be smaller for larger perturbations. We therefore analyse the large perturbation case as well.

\begin{figure}[t]
\centering
\includegraphics[trim=0 0 0 40, clip, width=\linewidth]{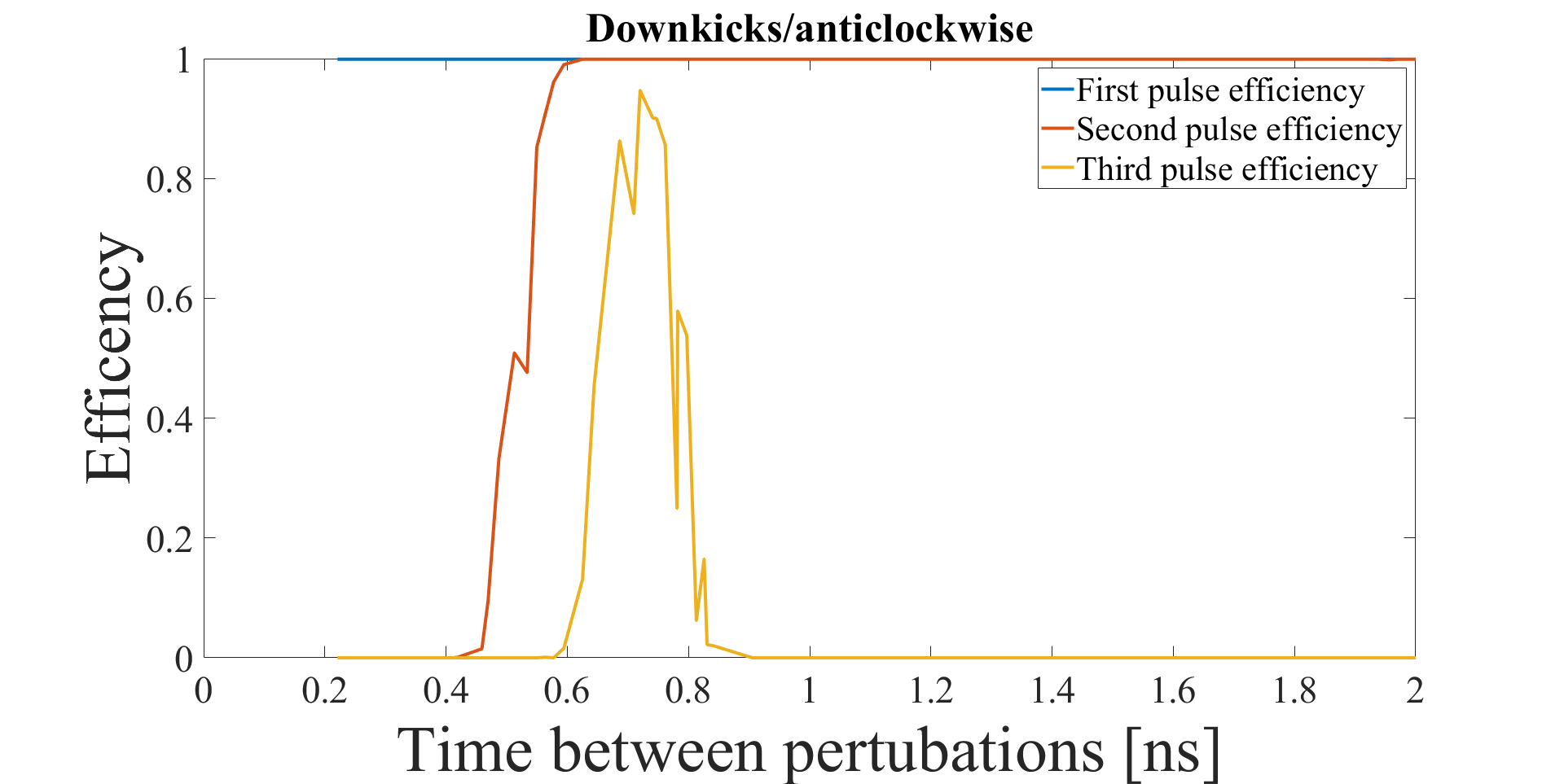}
\caption{The blue line shows the efficiency of the first perturbation. It always excites a pulse for times greater than 1~ns. The red line shows the efficiency of the second perturbation. Between 0.56~ns and 0.90~ns two perturbations can produce three pulses with the associated efficiency curve plotted in yellow.}
\centering
\label{down_effs}
\end{figure}

The large, 5.5~rad anticlockwise perturbations produce similar results to those seen for smaller perturbation amplitudes. The first perturbation always excites a pulse as shown in Figure \ref{down_effs} but the absolute refractory time is shorter, at 0.42~ns. After 0.63~ns the second perturbation is 100\% successful in triggering a pulse, as shown in Figure \ref{down_effs}. 

\begin{figure}[t]
\centering
\includegraphics[width=\linewidth]{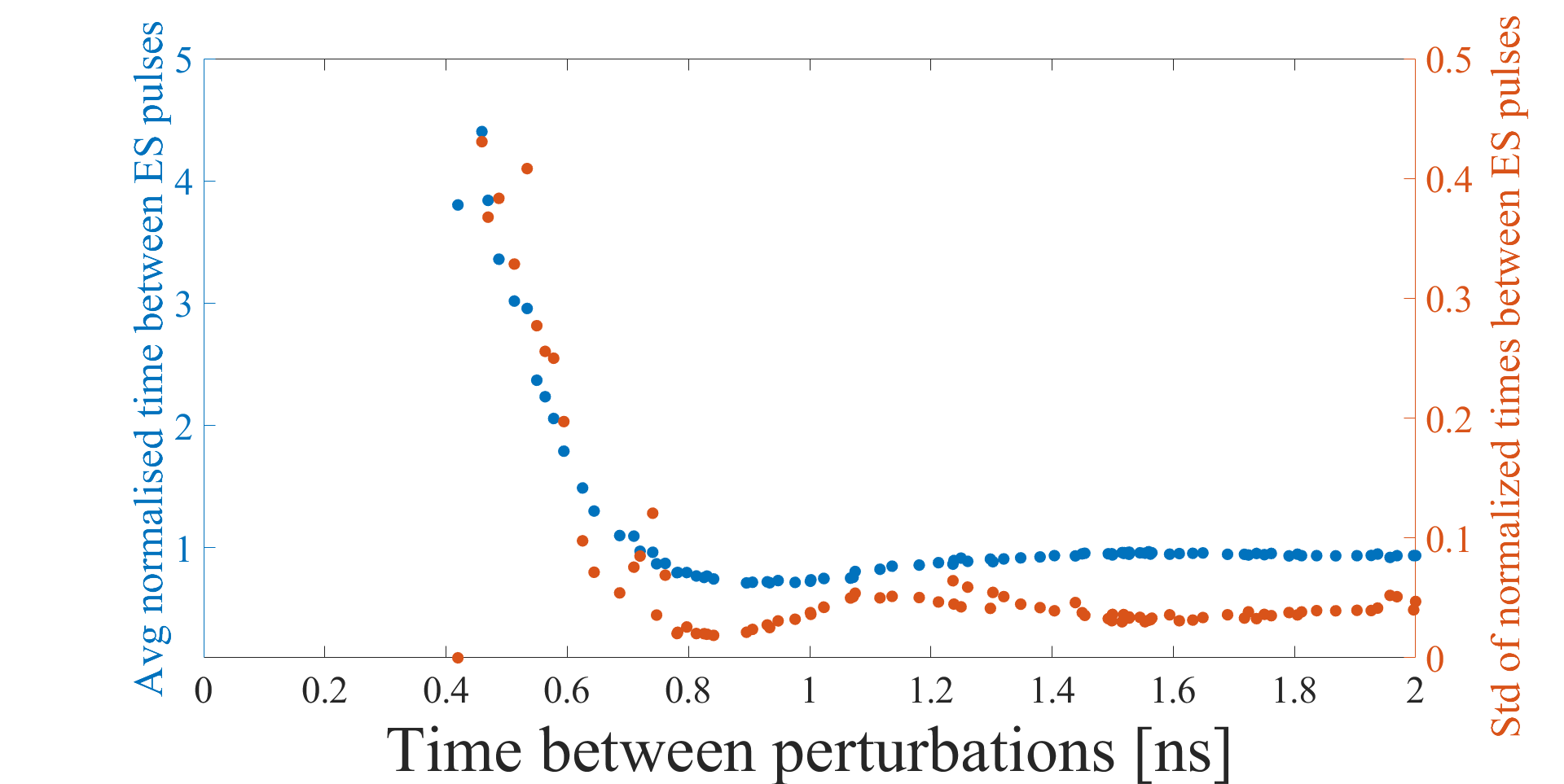}
\caption{The blue curve shows the average time between two ES pulses after two anti-clockwise perturbations are applied. The red curve shows the standard deviation of the times between two ES pulses.}
\centering
\label{Response_both}
\end{figure}

Figure \ref{Response_both} shows the average normalised time between ES pulses after two large anticlockwise perturbations were applied. Again, as with the smaller perturbation case, there is a large (and similar) variation in behaviour from the absolute refractory time up to the relative refractory time of approximately 1.4~ns. Thus, the overall behaviour is similar to the smaller perturbation, but as expected, the two refractory times are reduced.

\begin{figure}[t]
\centering
\includegraphics[width=\linewidth]{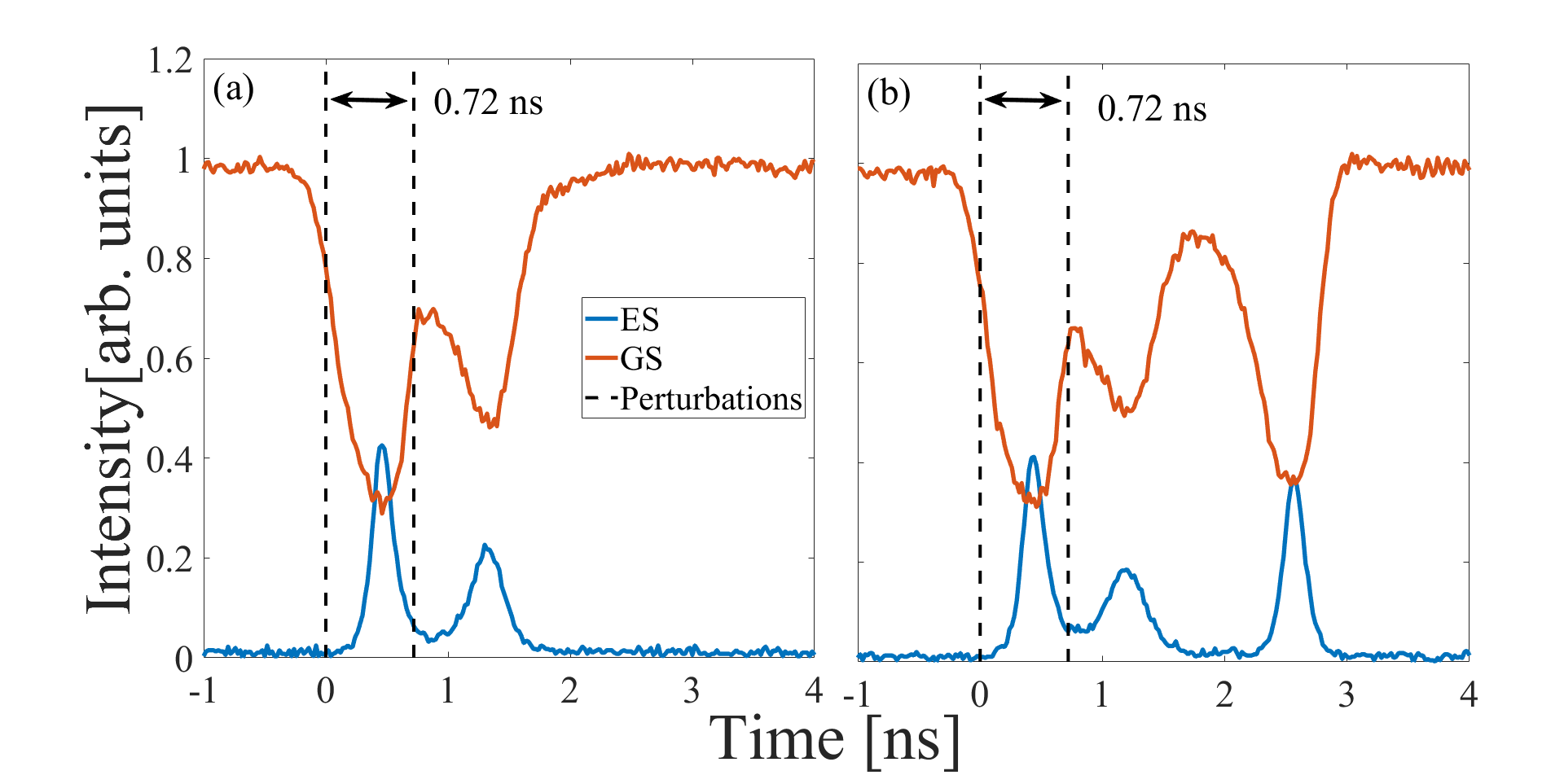}
\caption{(a) shows a timetrace with two pulses arising from two perturbations. There is a 95\% probability of three pulses occurring for this perturbation separation. (b) shows a timetrace with three pulses arising from two perturbations. In both (a) and (b) the time between the two perturbations is 0.72~ns.}
\centering
\label{triple_pulse}
\end{figure}

An interesting region is found when the second perturbation arrives between 0.56~ns and 0.90~ns after the first. In this region, a double pulse is sometimes found following the second perturbation. As usual the first perturbation always triggers a pulse, but when the second perturbation interrupts the first dropout, it can force the GS into a much shallower dropout with a corresponding small increase in ES intensity. A third dropout and full ES pulse can then fire after some delay as shown in Figure \ref{triple_pulse} (b). The efficiency curve for three ES maxima has a peak probability of 95\% where the perturbations are separated by 0.72~ns. On the occasions when three pulses do not fire, the two perturbations operate as conventionally expected and produce two GS dropouts with two ES pulses. Figure \ref{triple_pulse} (a) shows a timetrace where only two pulses fire for the same perturbation spacing as Figure \ref{triple_pulse} (b), indicating the sensitivity to noise.

\section{Conclusion}
We have measured both the absolute and relative refractory times for the asymmetric dual excitability obtained with optically injected QD lasers. For the anticlockwise case, the absolute refractory time is approximately 0.6~ns for small perturbations and is slightly lower, at 0.4~ns for large perturbations. The operationally important time - the relative refractory time - is approximately 1.5~ns for small perturbations and slightly reduced to 1.4~ns for large perturbations. For clockwise perturbations the absolute refractory time is approximately 1.02~ns while the relative refractory time is 1.25~ns. We acknowledge again that there is some ambiguity regarding the absolute refractory time for the clockwise case since there is an inhibition mechanism at low separation times. Nonetheless, 1.02~ns must elapse in order to excite a second pulse and so we pragmatically refer to it as the absolute refractory time. However, there is no ambiguity with regard to the relative refractory time.

These short refractory times for pulse generation at GHz order rates, up to $10^5$ times faster than biological neurons with pulses that are on the order of $10^6$ times shorter. This strongly suggests that such artificial neurons could be of use in systems where ultrafast analog computation is desired.

\end{document}